\documentstyle[12pt,epsf]{article}

\newcommand{\be}{\begin{equation}}
\newcommand{\ee}{\end{equation}}
\newcommand{\bea}{\begin{eqnarray}}
\newcommand{\eea}{\end{eqnarray}}

\begin{document}
\begin{titlepage}


\vspace{1in}

\begin{center}
\Large
{\bf Inflation, braneworlds and quintessence}

\vspace{1in}

\normalsize

\large{Greg Huey$^1$ \& James E. Lidsey$^2$}

\normalsize
\vspace{.7in}

{\em Astronomy Unit, School of Mathematical 
Sciences,  \\ Queen Mary, University of London, 
Mile End Road, LONDON, E1 4NS, U.K.}

\end{center}

\vspace{1in}

\baselineskip=24pt
\begin{abstract}
\noindent Inflationary cosmology is developed 
in the second Randall--Sundrum braneworld
scenario, where the accelerated 
expansion arises through potentials 
that are too steep to drive inflation 
in conventional cosmology. 
A 
relationship between the scalar and tensor 
perturbation spectra is derived that is
independent of both the inflaton potential and 
the brane tension. 
It is found that a single field with an inverse power law 
potential can act as both the inflaton and the quintessence 
field for suitable values of the 
brane tension. 
\end{abstract}

PACS NUMBERS: 98.80.Cq

\vspace{.7in}

\noindent $^1$Electronic mail: G.Huey $@$ maths.qmw.ac.uk

\noindent $^2$Electronic mail: jel@maths.qmw.ac.uk
 
\end{titlepage}


\section{Introduction}

\setcounter{equation}{0}

Recent accurate measurements of the first 
acoustic peak in the power spectrum of 
cosmic microwave 
background (CMB) anisotropies provide strong support that 
the universe is close to  spatial flatness
\cite{boomerang}. This is 
consistent with the most basic prediction 
of the inflationary scenario, where the universe underwent an 
epoch of accelerated expansion at or above the 
electroweak scale. (For recent reviews, see, e.g., Refs. 
\cite{reviews,abney}). 

On the theoretical side, 
there is currently enormous 
interest in  
understanding the cosmological 
consequences of viewing 
our
observable universe as a domain wall
or `brane' 
embedded in 
a higher--dimensional spacetime (the bulk) 
\cite{branerefs,rsII,modify,modify1,dr,linear,others,mwbh,hl,cll}. 
This
provides a new environment 
for studies of the very early universe  
and it is clearly important to investigate 
its implications
for inflation. In particular, 
the second Randall--Sundrum scenario (RSII) 
provides a solvable framework for addressing such questions. 

In the standard, chaotic scenario based on Einstein 
gravity minimally coupled to a self--interacting scalar 
`inflaton' field, 
$\phi$, inflation proceeds 
when $\alpha^2 \ll 2$ and ends when $\alpha^2 \approx 2$, 
where $\alpha^2 \equiv 
(m^2_4/8\pi)(d\ln V /d\phi)^2$
defines the slow--roll parameter, 
$m_4$ is the four--dimensional Planck mass and
$V(\phi )$ is the self--interaction potential 
of the field.  
However, conventional inflationary 
models must be  
fine--tuned if the potential 
is to be sufficiently flat. 
On the other hand, in the RSII braneworld
the expansion rate of the universe 
differs at high energies from that predicted by 
Einstein gravity 
in such a way that the 
friction acting on the scalar field is enhanced. 
Since the standard Hubble law 
must be recovered prior to the 
nucleosynthesis scale $(T \approx 
1{\rm MeV})$, a natural exit from inflation
ensues as the field accelerates down 
its potential \cite{cll}. 

One attractive feature of this scenario 
is that reheating  
arises naturally even when the potential 
does not have a global minimum. 
During any inflationary expansion, 
radiation is created
via gravitational 
particle production \cite{gpp}.
Although the density of radiation at the end of inflation is sub--dominant, 
the steep nature of the potential implies 
that the scalar 
field soon becomes kinetic energy dominated after 
inflation has ended \cite{kination,fj}. 
Consequently, the 
radiation can in principle 
dominate the universe at late times. 

Furthermore, since
the inflaton need not 
necessarily decay in this scenario,  
it may survive through
to the present epoch \cite{pv}.  
Depending on the form of its 
potential, therefore, 
it may also play the
role of the `quintessence' field
that 
has been invoked to explain the 
recent high--redshift supernovae observations
\cite{caldwell,snI}. 
These indicate  
that our universe is 
presently
entering a second accelerated 
expansion \cite{snI}. 

In this paper, 
we discuss 
inflation in the 
second Randall--Sundrum braneworld
and investigate 
the scalar and tensor perturbations 
produced during inflation in Section II \cite{rsII}. We 
proceed in Section III to 
consider inflation driven by a 
power law potential, $V \propto \phi^{-n}$ $(n>0)$, 
and determine the region of parameter space where  
the scalar field can act as 
the inflaton. We further determine the subregion 
of this space where the field can also act as quintessence. 
We conclude in Section IV with a discussion. 

\section{Inflation in the Randall--Sundrum II Model}

\setcounter{equation}{0}

\subsection{Dynamics}

In the second Randall--Sundrum (RSII) model,
a single, positive tension brane carrying the standard 
model fields 
is embedded in five--dimensional 
Einstein gravity with a negative (bulk) cosmological 
constant, $\Lambda$, and an infinite fifth dimension \cite{rsII}. 
In general, the backreactions 
between the bulk and the brane 
modify the field equations governing 
the brane's expansion \cite{modify,modify1,dr}. (For a review, 
see, e.g., Ref. \cite{maartens}).  
If the cosmological constant and brane tension, $\lambda$, 
are related by $\lambda^2=-6\Lambda m_5^6$, 
where $m_5^6= 4\pi \lambda m^2_4/3$ is the five--dimensional Planck mass, 
a projection of the higher--dimensional metric onto the brane 
world--volume results in a generalized Friedmann 
equation \cite{modify,linear}
\begin{equation}
\label{branefriedmann}
H^2 = \frac{8\pi}{3m_4^2} \rho \left[ 1+\frac{\rho}{2\lambda} 
\right]
\end{equation}
where $H \equiv \dot{a}/a$ denotes the Hubble parameter,
$\rho$ represents the matter confined to the brane, 
and a dot denotes 
differentiation with respect to cosmic time\footnote{There is also 
a `dark radiation' term arising from a
non--vanishing bulk Weyl tensor 
\cite{dr}. 
During inflation, however, such a term 
rapidly redshifts to zero and we consistently
neglect 
its effects in this paper.}. 
We assume that during inflation, a 
single 
scalar field that is confined to the brane
dominates the brane dynamics. 
Conservation of 
energy--momentum then implies that
\begin{equation}
\label{branescalar}
\ddot{\phi} +3H \dot{\phi} +V'=0
\end{equation}
where a prime denotes $d/d\phi$.

Eq. (\ref{branefriedmann}) determines the 
expansion rate of the brane's world--volume.
The quadratic correction implies that the universe 
expands at a faster rate for $\rho \gg \lambda$, 
leading to an enhanced friction on the 
scalar field \cite{mwbh,hl,cll}. This modification 
on the scalar field dynamics
can be quantified by 
defining the variables  
\begin{equation}
x \equiv \frac{\dot{\phi}}{\sqrt{2\rho}} , \qquad y \equiv 
\sqrt{\frac{V}{\rho}}
\end{equation}
Eqs. (\ref{branefriedmann}) and (\ref{branescalar}) may 
then be written  
as a plane--autonomous system \cite{gregreza}:
\begin{eqnarray}
\frac{dx}{dN} =-3x +3x^3 - \sqrt{\frac{3}{2}}\tilde{\alpha}
y^2  \nonumber \\
\frac{dy}{dN} = \sqrt{\frac{3}{2}} \tilde{\alpha} xy
+3yx^2 
\end{eqnarray}
where $N \equiv \ln a$, 
the variables are constrained
to lie on the unit circle, 
$x^2+y^2 =1$, and 
\begin{equation}
\label{effalpha}
\tilde{\alpha}^2 \equiv 
\frac{2\lambda}{\rho +2\lambda} \alpha^2 , \qquad \alpha^2 
= \frac{m^2_4}{8\pi} \frac{V'^2}{V^2}
\end{equation}
defines an effective slow--roll parameter\footnote{The sign of 
$\tilde{\alpha}$ is chosen to correspond to the sign of $dV/d\phi$.}. The 
advantage of expressing the field equations in 
this way
is that they are 
formally {\em identical} to those of the 
standard scenario \cite{clw} with the exception that the effective 
slow--roll parameter has acquired a correction given by 
$[1+(\rho/2\lambda ) ]^{-1}$. Thus, 
for a given $\alpha^2$, 
the field behaves as if the logarithmic derivative 
of its potential has been {\em reduced}.  
Hence, for $\rho \gg \lambda$, there may be inflation 
even though $\alpha^2  \gg 2$. 
If a given potential\footnote{We 
assume implicitly that the potential does not 
exhibit any special features.} satisfies 
such a condition,  
inflation will end when 
the inflaton's energy density 
still dominates the brane tension, 
$\rho \gg \lambda$. 

The continuous creation and redshifting of Hawking radiation 
in a time--varying gravitational field 
during inflation implies 
that at the end of inflation this radiation has 
a density $\rho_{R,e} \approx g_{\rm prod}H^4_e/100$
\cite{gpp,kination}, 
where $10 \le g_{\rm prod} \le 100$
denotes the number of particle species produced
at this epoch and
a subscript `e' denotes values at the end of 
inflation. 
The end of inflation is 
characterized by the 
condition $\tilde{\alpha}^2 \approx 1$:
\begin{equation}
\label{inflationend}
\frac{m^2_4}{4\pi} \frac{\lambda V'^2_e}{V^3_e} \approx 1
\end{equation}
and at this time, 
$\rho_{\phi , e} \approx V_e \gg \rho_{R, e}$. 
However, 
shortly after inflation, 
the scalar field rapidly becomes dominated
by its kinetic energy 
and behaves as a stiff perfect fluid, $\rho_{\phi} \propto 
a^{-6}$, whereas the radiation
scales as $\rho{_R} \propto a^{-4}$. 
If kinetic energy domination follows 
immediately after inflation (the `sudden--change' 
approximation), 
the radiation begins 
to dominate 
after the universe has expanded by a factor 
$(\rho_{\phi,e}/\rho_{R,e} )^{1/2}$. 
It then follows that the 
radiation's energy density 
at the moment 
when it begins to dominate 
is 
\begin{equation}
\label{raddom}
\rho_{R,{\rm dom}} =\frac{\rho^3_{R,e}}{\rho^2_{\phi ,e}} 
\approx 0.005 g^3_{\rm prod} \left( \frac{V_e}{\lambda} \right)^{10} 
\left( \frac{\lambda}{m^4_4} \right)^4 m^4_4
\end{equation}
where we further assume that 
the number of particle species does not change significantly 
between the end of inflation and radiation domination. 

There are a number of constraints that 
any inflationary model of this nature must  
satisfy in order to produce
a viable cosmology. The 
primary constraints 
are: (i) sufficient inflation 
must occur to solve the horizon 
and flatness problems; 
(b) the amplitudes 
of scalar and tensor 
perturbations produced during inflation must be consistent with 
microwave background anisotropies; (c) the spectral index 
of scalar perturbations must be sufficiently 
close to scale invariance;
(d) the 
standard Hubble expansion law must be recovered
and the universe must become radiation dominated 
before the onset of primordial nucleosynthesis; (e) 
the fraction of the universe's energy density in the form of 
the scalar field at that time must be sufficiently low;  
(f) if the scalar 
field has a late--time 
inflationary attractor (as is the case for inverse power law 
potentials, for example), then 
this second epoch of inflationary expansion 
must 
begin 
sufficiently late for 
large--scale structure 
to form.

The above is not
intended as an exhaustive list. 
Instead, it 
represents 
a necessary step that any successful model 
must overcome.
In principle, further constraints
could be imposed by 
considering higher-order 
effects, such as the influence of
scalar field inhomogeneities 
on structure formation 
and the CMB 
power spectrum.
Furthermore, if the temperature of the radiation 
after inflation exceeds $\sim 10^9$ GeV, the thermal production 
of gravitinos may lead to difficulties at nucleosynthesis
\cite{gravitino}.

\subsection{Perturbations}

The amplitudes of scalar and tensor perturbations 
produced during RSII inflation 
are given by \cite{mwbh,lmw}
\begin{eqnarray}
\label{scalaramp}
A^2_S = \frac{512\pi}{75m^6_4} \frac{V^3}{V'^2}
\left( 1+\frac{V}{2\lambda} \right)^3 \\
\label{tensoramp}
A_T^2 = \frac{4}{25 \pi m^2_4} H^2 F^2 (H/\mu )
\end{eqnarray}
respectively, where
\begin{equation}
\label{Fdef}
F(x) \equiv \left[ \sqrt{1+x^2} -x^2 {\rm sinh}^{-1} \left( 
\frac{1}{x} \right) \right]^{-1/2}
\end{equation}
and
\begin{equation}
\label{defx}
x \equiv \left( \frac{3}{4\pi \lambda} \right)^{1/2} H m_4
\end{equation}
and the normalization of Ref. \cite{abney} is 
chosen. In the low--energy limit, 
$\rho \ll \lambda$ $(x \ll 1)$, $F \approx 1$, whereas 
$F^2 \approx [27H^2m_4^2/(16\pi \lambda )]^{1/2}$ 
in the high--energy limit. 
The right--hand sides of Eqs. (\ref{scalaramp}) 
and (\ref{tensoramp}) are 
evaluated when the mode with comoving wavenumber, $k$, 
leaves the Hubble radius, $k=aH$ \cite{reviews,abney}. 
This implies that 
\begin{equation}
\label{kdef}
k(\phi) = a_e H(\phi ) \exp [ -N(\phi ) ]
\end{equation}
where $a_e$ denotes the 
value of the scale factor at the end of inflation 
and $N$ 
represents 
the number of e--foldings between a scalar field 
value, $\phi$, and the end of inflation, $\phi_e$:
\begin{equation}
\label{efold}
N \approx -\frac{8\pi}{m_4^2} \int^{\phi_e}_{\phi}
\frac{V}{V'} \left( 1+\frac{V}{2\lambda} \right) d\phi
\end{equation}

The corresponding 
spectral indices are defined by 
$n_S\equiv 1 + d \ln A^2_S/d \ln k$ and 
$n_T \equiv d\ln A^2_T /d \ln k$. 
Differentiating Eq. (\ref{tensoramp}) 
with respect to comoving wavenumber and 
substituting in Eqs. (\ref{Fdef}), 
(\ref{defx}) and 
(\ref{kdef}) implies that 
\begin{equation}
\label{nt}
n_T= -2 \frac{1}{N'}\frac{x'}{x} \frac{F^2}{\sqrt{1+x^2}}
\end{equation}
where the slow--roll approximation, 
$\tilde{\alpha}^2 \ll 1$, 
has been implicitly assumed. Further substitution of Eqs. 
(\ref{branefriedmann}), (\ref{tensoramp}) 
and (\ref{efold}) into Eq. (\ref{nt}) then eliminates 
any direct dependence on the inflaton 
potential and yields a 
`consistency' equation
\begin{equation}
\label{consistency}
\frac{A^2_T}{A^2_S} =  -\frac{1}{2} n_T
\end{equation}
relating the ratio of the tensor and 
scalar amplitudes with the tensor spectral index. 
Remarkably, Eq. (\ref{consistency}) 
is {\em independent}
of the brane tension, $\lambda$, and 
consequently 
has precisely the same form as the 
corresponding consistency equation arising 
in the standard, chaotic inflationary scenario \cite{abney}. 

Finally, 
the spectral indices 
in the high--energy limit 
are given 
in terms of the potential and its derivatives 
by
\begin{eqnarray}
\label{scalarindex}
n_S-1 \approx -\frac{m^2_4\lambda}{2\pi V}\left[ 3 
\frac{V'^2}{V^2} -\frac{V''}{V} \right] \\
\label{tensorindex}
n_T \approx -\frac{3m^2_4}{4\pi} \frac{\lambda V'^2}{V^3}
\end{eqnarray}
During inflation, the 
logarithmic slope, 
$\alpha^2$, of the potential is 
large and in general this produces 
spectra that 
differ
significantly from the 
scale invariant form $(n_S=1, n_T=0)$.
For example, 
models driven by an exponential inflaton potential 
predict that $n \approx 0.92$ for all 
values of $\alpha^2$ \cite{cll}.
Moreover,  
Eq. (\ref{consistency}) implies that 
a strong tilt in the tensor
spectrum increases the 
relative amplitude of 
the gravitational waves. 
These features represent 
important signatures and provide strong constraints on 
this class 
of model. 

It is also worth noting that 
the scalar perturbations vary as $A_S 
\propto V^2\lambda^{-3/2}\alpha^{-1}$. 
It might be expected, therefore, 
that since $\alpha^2$ 
is large in these models, 
the last 50 e--foldings of inflation could 
occur at a higher energy than in the standard scenario
without violating the 
COBE normalization constraint, $A_S (N\approx 50) =2\times 10^{-5}$
\cite{nor}.
However, a larger $\alpha^2$ 
implies that the magnitude
of the potential $N$ e--foldings
before the end of
inflation may also be larger. 
For the  exponential potential, 
$V(N) \approx \lambda \alpha^2 N$
and it is this latter effect that dominates 
\cite{cll}. 
This is also the case for the power law models 
we consider in the following Section. 
Thus, increasing $\alpha^2$ may actually increase
the amplitude of the perturbations laid down 
at $N \approx 50$. This implies that 
inflation must end at a lower energy scale 
if acceptable density perturbations are 
to be produced and 
this limits 
not only the radiation density 
at the end of inflation, but 
also
results in a smaller expansion 
factor from the 
end of inflation to the 
nucleosynthesis epoch. 
Depending on the particular form of the potential, 
this may lead to a significant constraint on the model. 

\section{Inverse power law models}

\setcounter{equation}{0}

In this Section, we consider the class of power law potentials
\begin{equation}
\label{powerlaw}
V= \frac{M^{4+n}}{\phi^n}
\end{equation}
where $\{ M , n\}$ are constants. 
The corresponding model with an exponential potential 
was considered in Ref. \cite{cll}. 
Inverse power law models
are interesting for a number of reasons.
In conventional cosmology, 
they drive `intermediate' inflation 
\cite{barrow} and 
typically produce significant tensor perturbations 
for almost scale--invariant scalar fluctuations \cite{barrowlid}. 
They arise in supersymmetric condensate models of QCD
\cite{qcd}
and 
can in principle 
act as a source of 
quintessence \cite{pebratra,balbi}.

In standard cosmology, inflation does not
begin until 
the scalar field exceeds a critical value, 
$\phi^2_{\rm crit} = n^2m^2_4/(16\pi )$. However, 
the slow--roll 
parameter in the 
RSII braneworld model is 
\begin{equation}
\tilde{\alpha}^2 \approx \frac{n^2 
\lambda m^2_4}{4\pi M^{4+n}} \phi^{n-2}
\end{equation}
and inflation is therefore possible 
for $n> 2$ if $\phi < \phi_e$, where
\begin{equation}
\phi^{n-2}_e \approx \frac{4\pi M^{4+n}}{n^2 \lambda m^2_4}
\end{equation}
It follows, therefore, that 
when the brane tension exceeds 
\begin{equation}
\label{twophases}
\lambda > \frac{4^{n-1}\pi^{n/2}}{n^n} \left( \frac{M}{m_4} \right)^{n+4}
m^4_4
\end{equation}
the universe inflates for $\phi < \phi_e$, 
undergoes subluminal expansion for $\phi_e < \phi <\phi_{\rm crit}$ 
and enters a second phase of inflation for $\phi
> \phi_{\rm crit}$. 

We now consider the constraints on the model
when Eq. (\ref{twophases}) is satisfied. 
Corrections 
to Newtonian gravity are negligible above 
$1 {\rm mm}$ scales if 
$\lambda^{1/4} > 100 {\rm GeV}$ \cite{mwbh}. This ensures that 
the standard Hubble law is recovered prior to 
nucleosynthesis. 
The value of the slow--roll parameter $N$ e--foldings 
before the end of inflation is 
\begin{equation}
\label{slowN}
\tilde{\alpha}^2_N \approx \frac{n}{n+(n-2)N}
\end{equation}
and this implies that the scalar spectral index is given by 
\begin{equation}
\label{scalartilt}
n_S \approx 1-\frac{4n-2}{n+(n-2)N}
\end{equation}
Eq. (\ref{scalartilt}) is independent 
of the mass parameter, $M$, and the brane tension, $\lambda$. 
Conventionally, observations 
constrain the spectral tilt at $N \approx 50$. 
For fixed $N$, the index is only weakly 
dependent on the power, $n$, 
and is bounded from above 
by $n_S< 1- 4/(1+N) \approx 0.92$ as $n \rightarrow \infty$. 
It is interesting that this 
upper limit corresponds precisely to the value 
produced by an exponential potential \cite{cll}. 
This value is consistent with observations
\cite{boomerang}, but could be ruled out in the near future. 
An observationally acceptable lower limit
of $n_S \approx 0.9$  
on the spectral index corresponds  
to a limit of $n >7$ on the power.
The ratio of scalar to tensor amplitudes in this model 
is given by
\begin{equation}
\frac{A^2_T}{A^2_S} = \frac{3n}{2[n+(n-2)N]}
\end{equation}
and also recovers the exponential 
potential value as 
$n \rightarrow \infty$. It is anticipated that 
the Planck satellite will be sensitive to 
gravitational waves with such an 
amplitude. 

When establishing the constraints due to the COBE 
normalization, $A_S =2 \times 10^{-5}$ \cite{nor}, 
it proves convenient to 
define the dimensionless parameters
\begin{equation}
\label{YZ}
Y \equiv (8 \pi)^2 \frac{\lambda}{m^4_4}, \qquad Z \equiv 
\frac{n^n}{(8\pi)^{n/2}} \left( \frac{\lambda}{m^4_4} 
\right) \left( \frac{M}{m_4} \right)^{-(n+4)}
\end{equation}
Substituting Eq. (\ref{YZ}) 
into Eq. (\ref{scalaramp}) implies that 
\begin{equation}
\label{cobeconstraint}
YZ^{6/(n-2)} \left[ 2 + \frac{2(n-2)N}{n}
\right]^{(4n-2)/(n-2)} \approx 2.4 \pi^2 \times 10^{-7}
\end{equation}
This constraint relates the brane tension, $\lambda$, 
to the inflaton's mass parameter, $M$. We employ it to 
eliminate $Z$ in favour of expressing the allowed region of 
parameter space in terms of $\{Y,n \}$. 

To be consistent, we must 
verify that at least 50 e--foldings 
of inflation are possible in this model. Quantum 
gravitational effects are expected to become 
important when the inflaton's energy density exceeds the 
five--dimensional Planck scale 
and
the 
assumption that the inflaton field is confined 
to be brane may not be valid above this scale.
We therefore require that 
$V <m^4_5 \approx \lambda^{2/3} m^{4/3}_4$
and this is equivalent to 
\begin{equation}
\label{quantgrav}
V< 10^{-2} Y^{2/3}
\end{equation}
>From the definition 
(\ref{YZ}), it follows that 
\begin{equation}
\label{VasZ}
\frac{V}{\lambda}= Z^{2/(n-2)} \left( \frac{2}{\tilde{\alpha}^2}
\right)^{n/(n-2)}
\end{equation}
and substituting the constraint (\ref{cobeconstraint}) 
into Eq. (\ref{VasZ}) then yields a relationship between the
parameter, $Y$, 
and the  
magnitude of the potential $N=50$ e--foldings 
before the end of inflation: 
\begin{equation}
\label{VasY}
\frac{V_N}{m^4_4} \approx \frac{(600\pi^2)^{1/3}}{8\pi^2} \left[ 
\frac{2+(n-2)N}{n} \right]^{-1/3} Y^{2/3} A^{2/3}_{S,{\rm COBE}}
\end{equation}
Thus, the smallness of the perturbations 
on COBE scales ensures that condition (\ref{quantgrav}) 
is automatically satisfied and 
sufficient inflation is therefore possible. 

\begin{figure}
\begin{center}
\leavevmode\epsfysize=5.5cm \epsfbox{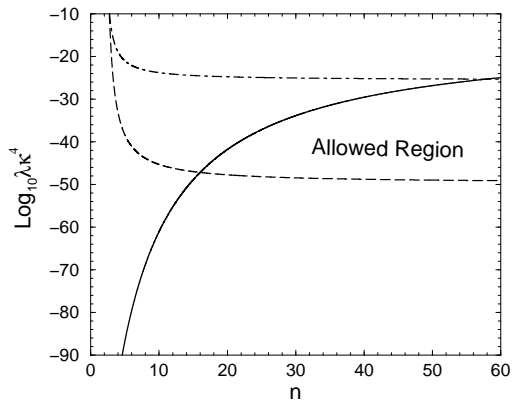}\\ 
\end{center}
\caption{\small The parameter space 
where a field with a power law potential 
(\ref{powerlaw}) can lead to successful braneworld 
inflation in the second Randall--Sundrum 
scenario. The allowed region is bounded by the solid and dashed lines, 
corresponding to the late--time inflation constraint (\ref{latetime}) 
and the radiation--domination constraint (\ref{raddompower}), 
respectively. The 
radiation temperature immediately after inflation 
exceeds  $10^9$ GeV above the dot--dashed line. 
The details of the constraints are discussed in the 
text and $\kappa^2 \equiv 8 \pi /m^{-2}_4$.} 
\label{constraintplot}
\end{figure}

We must now consider the question of reheating in this 
scenario. 
Radiation domination is achieved before nucleosynthesis 
if $\rho_{R, {\rm dom}} > 1.6\times 10^{-88} m^4_4$,  
where $\rho_{R, {\rm dom}}$ can be estimated analytically 
if the sudden--change approximation of 
Eq. (\ref{raddom}) is employed. In this case, 
we find that
\begin{equation}
\label{raddompower}
\rho_{R, {\rm dom}} \sim 16\left( 
\frac{g_{\rm prod}}{900} \right)^{3} \left( 2Z \right)^{10}
\left( Y \right)^{4}
\gg 1.6 \times 10^{-88} m^4_4 
\end{equation}
However, in reality 
the scalar field takes a finite amount of time 
to complete the transition from potential--energy to 
kinetic--energy 
domination and 
the sudden--change approximation therefore 
tends to overestimate the 
temperature of the universe when the radiation begins to 
dominate. 
Since this correction 
is difficult to 
calculate analytically, we 
have determined the transition epoch numerically. 
For the parameter values of interest, we find that 
the sudden--change
approximation yields a value 
for the scalar field density  
that is typically a 
factor of $100$ too low, although 
this factor varies
by less than 10\% over the allowed range of parameter values, 
out to $n \sim 50$. The numerical result is shown 
in Fig. \ref{constraintplot}.  

At the nucleosynthesis epoch, 
the expansion rate is constrained by light 
element abundances  such 
that the density parameter of the scalar field 
is bounded by 
$\Omega _{\phi } \leq 0.2$
\cite{fj}. 
We do not 
employ this constraint
directly  
in determining the allowed region 
of parameter space $(n,Y,Z)$
because it is difficult to express it in 
terms of these variables. Instead, we first 
impose the 
other constraints and then verify that 
this condition is satisfied over the corresponding 
region of interest. 

Finally, we must ensure that the second phase 
of inflation begins when the universe
is sufficiently old. 
Determining when this occurs 
is a complicated  question in general, 
because it depends on several dynamical 
factors, in particular on $\dot{\phi}^2$, $\Omega_{\phi}$
and $\alpha^2$. However, 
if 
late--time inflation arises when the scalar field 
is at its attractor value, 
it follows that 
inflation begins at approximately the same time that 
the condition 
$V'' \approx H^2$ breaks down. In this case, 
the condition that inflation is just starting at the present epoch 
is given by  
$\alpha^2 \approx 2$, where $\alpha$, as defined in Eq. 
(\ref{effalpha}), 
is evaluated 
when $V'' \approx H_0^2$. 
Since 
\begin{equation}
\frac{V''}{m^2_4} = (8\pi )^{(n+2)/2} \left( 
\frac{n+1}{n^{n+1}} \right) \left( \frac{M}{m_4} 
\right)^{n+4} \alpha^{n+2}
\end{equation}
this yields the 
constraint 
\begin{equation}
\label{latetime}
Z > \frac{2^{(n-4)/2} (n+1)}{\pi n} Y \left( 
\frac{H_0}{m_4} \right)^{-2}
\end{equation}

The allowed region of parameter space is 
indicated in Fig. \ref{constraintplot}. 
The radiation dominates the scalar field before nucleosynthesis 
in the region above the dashed line and 
the secondary phase of inflation 
does not begin until after the 
present epoch in the region below the solid line. All other 
constraints are satisfied in this regime. 
Consequently, there 
is a lower limit of $n>15$ on the power of the potential.  
The radiation--domination constraint is 
weakly dependent 
on the brane tension above this value, 
implying that $\lambda^{1/4} > 10^{6}$ GeV. 
Below the dot--dashed line, the
temperature of the radiation at the end of inflation 
is less then 
$10^9$ GeV, corresponding to the critical temperature where 
thermal gravitino production may be problematic \cite{gravitino}. 
This constraint is satisfied for all $n <50$. 

We now consider the 
quintessence scenario for 
the inverse power law model (\ref{powerlaw}). 
Maeda has 
found that in the radiation--dominated 
RSII scenario, the fraction of the scalar 
field density decreases in the high--energy $(\rho \gg 
\lambda )$ regime for $n \ge 2$ and 
has found that 
quintessence is possible 
for $n> 4$ \cite{maeda}. We are interested in 
the possibility that such a field 
can act as both the inflaton and quintessence
fields. In this case, the COBE normalization (\ref{cobeconstraint}) 
restricts the allowed value of the potential parameter, $M$.

\begin{figure}
\begin{center}
\leavevmode\epsfysize=5.5cm \epsfbox{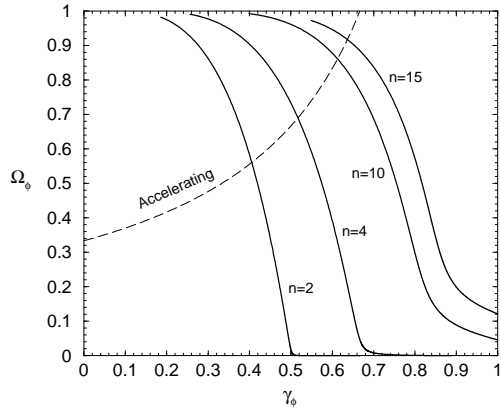}\\ 
\end{center}
\caption{\small  The phase plane illustrating the 
density parameter of the scalar field, 
$\Omega_{\phi}$, 
as a function of its
equation of state, $\gamma_{\phi} = (\rho +p)/ \rho$.
The universe is accelerating above the dashed line.} 
\label{phaseplane}
\end{figure}

At low energies, 
the late--time attractor 
is a `tracking' solution \cite{tracking}:
\begin{equation}
\label{tracking}
\Omega_{\phi} =\frac{3\gamma_{\phi}}{\alpha^2}, \qquad 
\gamma_{\phi} = \frac{n\gamma_{\rm bgrd}}{n+2}
\end{equation}
for $\alpha^2 > 3\gamma_{\rm bgrd}$, where 
the barotropic index of the scalar field 
is defined by $\gamma_{\phi} \equiv (\rho +p) /\rho$
and $\gamma_{\rm bgrd}$ is the 
corresponding index for the 
background fluid. 
This leads to the 
de Sitter solution 
in the infinite future, where 
$\gamma_{\phi} \rightarrow 0$ and $\Omega_{\phi} \rightarrow 1$. 
Fig. \ref{phaseplane} 
illustrates the tracking evolution of the field 
in the $(\Omega_{\phi} , \gamma_{\phi} )$ 
parameter space for different values of $n$. 
Initial conditions on $( \Omega_{\phi} , \gamma_{\phi} , 
\alpha^2 )$ are chosen in such 
a way that the field is tracking
by the time of nucleosynthesis 
and its energy density
still satisfies $\Omega_{\phi} \ll 0.1$ at 
matter--radiation equality. Since $\Omega_{\phi}$ is monotonically 
increasing, this 
ensures that $\Omega_{\phi} < 0.15$ at nucleosynthesis. 

A necessary condition for successful 
quintessence is that the present--day universe must be above the 
dashed line, corresponding to accelerated
expansion. However, it is not possible to satisfy 
the condition $\Omega_{\phi} \approx 
0.7$ and $\gamma_{\phi} 
< 2/3$ 
for the range of $n$ consistent with 
Fig. \ref{phaseplane}. We must conclude, therefore, that  
the inflaton field cannot provide the missing dark energy 
in this braneworld model
if it has already reached its late--time attractor by the time 
of nucleosynthesis.

An alternative 
possibility is that the scalar field may overshoot its attractor 
value if its initial kinetic energy is sufficiently high 
\cite{tracking}. 
In this case, 
its energy density 
falls below the value corresponding to the tracking 
condition (\ref{tracking}). 
It then follows that 
the field must become
fixed at a point on its potential
in order to allow $\Omega_{\phi}$ 
to increase towards its tracking value. 
During this 
time, 
it behaves as a cosmological constant, 
with an equation of state very close to $\gamma_{\phi} 
\approx 0$. 

\begin{figure}
\begin{center}
\leavevmode\epsfysize=5.5cm \epsfbox{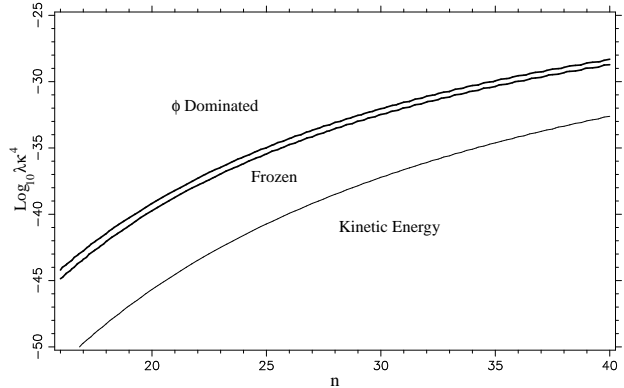}\\ 
\end{center}
\caption{\small 
Regions in the parameter space where the field at the present 
epoch is dominated by either its kinetic  
or potential energy. For each point in the 
$( \lambda , n )$ plane, initial conditions 
are chosen in the numerical calculation to correspond to the 
conditions at the end of successful inflation. The 
dependence on the coupling parameter, $M$, 
has been eliminated by imposing the COBE normalization constraint
(\ref{cobeconstraint}) and the ratio of scalar field 
energy density to that of ordinary matter is calculated assuming 
$\Omega_{\rm matter} h^2 
\approx 0.3 \times 0.65^2$. 
The four--dimensional 
Planck mass is $m^2_4 = 8\pi/\kappa^2$.}
\label{stop}
\end{figure}

Fig. \ref{stop} illustrates the region 
of the $(\lambda , n )$ plane consistent 
with this behaviour at the present epoch.
In the region marked `kinetic energy' 
the field is still 
rolling today and is dominated by its 
kinetic energy, although its density parameter 
is negligible, $\Omega_{\phi} \ll 1$.
In the region marked `frozen', the field 
has ceased rolling and is dominated by its 
potential ($\gamma_{\phi} \sim 0$). Finally, 
in the region denoted `$\phi$ dominated' 
the density of ordinary 
matter 
has become negligible, i.e., 
$\Omega_{\phi} \sim 1$ and $\gamma_{\phi} \approx 0$.
Between these latter two regions is a thin area bounded 
by the two solid lines which would produce
a density parameter $ 0.1 \le \Omega_{\phi} \le 0.9$ today. 
The upper line represents
$\Omega_{\phi} = 0.9$, and the lower $\Omega_{\phi} = 0.1$

\section{Discussion}

\setcounter{equation}{0}

In this paper, we have considered 
models of inflation in the second Randall--Sundrum 
braneworld scenario. 
It was found that the consistency equation 
(\ref{consistency}) 
relating the tensor and scalar perturbation spectra 
is {\em independent} of the brane tension
to lowest--order in the slow--roll 
approximation. 
This equation is given 
entirely in terms of observable parameters and is 
also independent
of the functional form of the potential driving inflation. 
In the event of a positive detection of a primordial gravitational 
wave background, confirmation of such a relationship would 
provide strong support for inflation. 
However, one could not determine from this equation alone 
whether inflation arose within 
the context of a braneworld 
scenario or from standard relativistic cosmology. 
In the latter case, next--order corrections 
in the slow--roll parameters lead to corrections involving 
the scalar spectral index and it would 
be interesting to determine whether similar terms 
arise in the braneworld model. This would 
involve extending the analyses of Refs. 
\cite{mwbh} and \cite{lmw}
along the lines discussed in Ref. \cite{sl93}. 

Furthermore, we have found that for the inverse power law potential, 
the 
scalar field can successfully drive braneworld inflation and, 
for appropriate choices of parameters, can act as 
the quintessence field at the present epoch. 
The spectral indices of the scalar and tensor 
perturbations are only weakly dependent on the power of the potential, 
$n$. 
A detectable tilt towards larger wavelengths 
is predicted for the scalar perturbations 
and the gravitational wave amplitude is within the 
range of sensitivity anticipated for the Planck satellite. 
In view of this, 
it is expected that models of this type should 
produce a CMB power spectrum where the first 
acoustic peak is relatively low. Although a  detailed analysis 
of the anisotropies produced is beyond the 
scope of the present work, such an investigation 
may provide further observational tests of these models
in light of the  
recent measurements of the first acoustic peak \cite{boomerang}.

\vspace{.7in}

{\bf Acknowledgments} GH is supported by the Particle Physics 
and Astronomy Research Council (PPARC). JEL is 
supported by the Royal Society. We thank A. Liddle for a 
helpful discussion.

\end{document}